\documentstyle[12pt]{article}
\begin{document}
\noindent
\begin{center}
{\Large {\bf Holographic Dark Energy Model \\and Scalar-Tensor
Theories\\}} \vspace{2cm}
 ${\bf Yousef~Bisabr}$\footnote{e-mail:~y-bisabr@srttu.edu.}\\
\vspace{.5cm} {\small{Department of Physics, Shahid Rajaee
University,
Lavizan, Tehran 16788, Iran.}}\\
\end{center}
\vspace{1cm}
\begin{abstract}
We study the holographic dark energy model in a generalized scalar
tensor theory.  In a universe filled with cold dark matter and
dark energy, the effect of potential of the scalar field is
investigated in the equation of state parameter.  We show that for
a various types of potentials, the equation of state parameter is
negative and transition from deceleration to acceleration
expansion of the universe is possible.

\end{abstract}
\vspace{3cm}
\section{Introduction}
It is strongly believed that our universe is in a phase
experiencing an accelerated expansion. The recent observations
regarding the luminosity-redshift relation of type Ia supernovae
\cite{sup} in association with observations on Cosmic Microwave
Background Radiation \cite{cm} demonstrate this cosmic
acceleration. The simplest candidate to produce this cosmic
speed-up is the cosmological constant, the energy density
associated with quantum vacuum.  However, there are at least two
problems for associating cosmic acceleration with the cosmological
constant.  Firstly, theoretical estimates on its value are many
order of magnitude larger than observations \cite{win}.  Secondly,
it is simply a constant, namely that it is not diluted with
expansion of the universe.  This latter is specifically important
in the sense that there are observational evidence \cite{R}
demonstrating that the cosmic expansion is a recent phenomena and
the universe must have passed through a deceleration phase in the
early stages of its evolution.  This deceleration phase is
important for successful nucleosynthesis as well as for the
structure formation. We therefore need a field evolving during
expansion of the universe in such a way that its dynamics makes
the deceleration parameter have a signature flip from  positive in
the early stages of matter dominated era to negative in the
present
stage.\\
This fact has motivated people to consider many dynamical models
trying to explain the cosmic acceleration.  Among these models
there are those that assumes a mysterious cosmic fluid with
sufficiently large and negative pressure, dubbed dark energy.
These models are usually invoked a scalar field which during its
evolution takes negative pressure by rolling down a proper
potential.  There is a large class of scalar field models in the
literature including, quintessence \cite{q}, k-essence \cite{k},
tachyons \cite{tach} , phantom field \cite{ph}, quintom \cite{qui}
, and so forth.  However, most of these models are not compelling
enough and require fine tuning of parameters to be consistent with
observations.\\
There is another proposal, first formulated by 't Hooft and
Susskind \cite{suss}, which recently has attracted much attention
as a possible solution to the dark energy problem.  The basic
idea, dubbed holographic principle, is that the number of degrees
of freedom of a physical system scales with its bounding area
rather than with its volume.  For an effective quantum field
theory in a box of size $L$ with an ultraviolet (UV) cutoff
$\Lambda$, the entropy $S$ scales extensively as $S\sim
L^{3}\Lambda^{3}$. However the peculiar thermodynamics of black
holes has led Bekenstein \cite{bek} to postulate that the maximum
entropy in a box of volume $L^{3}$ behaves non-extensively,
growing as the area of the box.  In this sense there is a
so-called Bekenstein entropy bound
\begin{equation}
S=L^{3}\Lambda^{3}\leq S_{BH}\equiv \pi L^{2}M^{2}_{p}
\label{0a}\end{equation} where $S_{BH}$ is the entropy of a black
hole of radius $L$, and $M_{p}\equiv (8\pi G)^{-\frac{1}{2}}$
stands for the reduced Planck mass.  It is important that in this
relation the length scale $L$ providing an Infrared (IR) cutoff is
determined by the UV cutoff $\Lambda$ and can not be chosen
independently. However such a non-extensive scaling law seems to
provide a breakdown of quantum field theory at large scales.  To
reconcile this breakdown with the success of local quantum field
theory in describing observed particle phenomenology, Cohen et al.
\cite{co} proposed a more restrictive bound.  Since the maximal
energy density in the effective theory is of the order
$\rho_{\Lambda}=\Lambda^{4}$, requiring that the energy in a given
volume not to exceed the energy of a black hole of the same size
results in the constraint
\begin{equation}
L^{3}\rho_{\Lambda}\leq L M^{2}_{p} \label{02}\end{equation} If we
take the largest value of the length scale $L$ as the IR cutoff
saturating the inequality (\ref{02}), we then obtain the
holographic dark energy density
\begin{equation}
\rho_{\Lambda}=3c^{2} M^{2}_{p}L^{-2} \label{03}\end{equation} in
which $3c^{2}$ is a numerical constant.  It is interesting to note
that if the length scale $L$ is characterized by the size of the
universe, the Hubble scale $H^{-1}$, then equation (\ref{03})
gives a vacuum energy density of the right order of magnitude
consistent with observations \cite{co}.  It is however pointed out
that this yields a wrong equation of state parameter for dark
energy, and other possible values for $L$ should be chosen such as
the size of the future event horizon \cite{li}.  This conclusion
is, however,  based on the assumption of an independent evolution
of energy densities of dark energy and dark matter.  It is shown
\cite{pp} that, if there is any interaction between these two
components the identification of $L$ with $H^{-1}$ is possible and
the equation of state parameter indicates late time acceleration.
In a recent work \cite{ban}, an interacting holographic dark
energy is studied in Brans-Dicke theory. It is shown that in this
framework there is a noninteracting limit and for a given set of
parameters the equation of state parameter can be negative. Here
we would like to generalize this work to the Brans-Dicke theory
with a self-interacting potential.  We shall assume that the
matter contained in the universe consists of cold dark matter and
an interacting holographic dark energy. We show that the potential
term improves the behavior of the set of parameters. In
particular, for a zero potential term the late time acceleration
constrains the evolution of the gravitational coupling in the
Brans-Dicke theory in such a way that it is only allowed to
increase with expansion.  It is certainly unnatural if dynamical
evolution of the scalar field would lead to this undesirable
situation of going through an infinitely strong gravitational
effect.  We shall show that this behavior is improved for a
nonzero potential.\\
This paper is organized as follows: In section 2, we first
consider late time acceleration based on holographic principle in
a generalized scalar-tensor theory. We assume that the matter
contained in the universe consists of cold dark matter and
holographic dark energy. Here these two types of matter are not
conserved separately due to an interaction. As special cases, we
then consider the limits of the model in general relativity and
Brans-Dicke theory with potential. In the first parametrization
(general relativity), we show that there is a non-interacting
limit in the spatially curved Robertson-Walker spacetime. In the
Brans-Dicke parametrization, we specifically study the effect of
the potential of the scalar field in the behavior of equation of
state parameter and the deceleration parameter.
In section 3, we offer some concluding remarks.\\\\

\section{The Model}
We consider a model in which gravity is described by a
scalar-tensor theory.  The most general action functional for
these theories is given by \footnote{Our sign convention is (-+++)
and we work in units in which $\hbar=c=1$.}
\begin{equation}
S=\frac{1}{2} \int d^{4}x \sqrt{-g}~
\{F(\phi)~R+U(\phi)~g^{\alpha\beta} \nabla_{\alpha}\phi
\nabla_{\beta}\phi+V(\phi)\}+ S_m(g_{\mu\nu})
\label{2a}\end{equation} where $S_m(g_{\mu\nu})$ is the matter
field action, $g$ is the determinant of the metric $g_{\mu\nu}$,
$R$ is the Ricci scalar and the functions $F(\phi)$, $U(\phi)$ and
$V(\phi)$ are arbitrary functions of the real scalar field $\phi$.
The only constraint on these functions is that $F(\phi)>0$
ensuring that graviton carries positive energy \cite{gra}.  We
also note that the matter action does not involve $\phi$ which
means that the whole theory respects the
weak equivalence principle \cite{will}.\\
Variations with respect to $g_{\mu\nu}$ and $\phi$ gives the field
equations
\begin{equation}
F(\phi)G_{\mu\nu}=(T_{\mu\nu}+
T^{\phi}_{\mu\nu})\label{2b}\end{equation}
\begin{equation}
2U(\phi)\Box\phi+2\nabla_{\gamma}U(\phi)\nabla^{\gamma}\phi-\frac{dU}{d\phi}
\nabla_{\gamma}\phi\nabla^{\gamma}\phi+\frac{dF}{d\phi}R-\frac{dV}{d\phi}=0\label{2c}
\end{equation}
where
\begin{equation}
T^{\phi}_{\mu\nu}=U(\phi)(\nabla_{\mu}\phi \nabla_{\nu}\phi
-\frac{1}{2}g_{\mu\nu}\nabla^{\alpha}\phi
\nabla_{\alpha}\phi)+(\nabla_{\mu}\nabla_{\nu}-g_{\mu\nu}\Box)F(\phi)
-\frac{1}{2}g_{\mu\nu}V(\phi) \label{2d}\end{equation}
\begin{equation}
T_{\mu\nu}=-\frac{2}{\sqrt{-g}}\frac{\delta S_{m}}{\delta
g^{\mu\nu}} \label{2e}\end{equation} In a cosmological context, we
take $T_{\mu\nu}$ to consists of two interacting components, a
pressureless dark matter and a holographic dark energy. Since both
components do not evolve independently, a source (or loss) term
must enter their energy balances
\begin{equation}
\dot{\rho_{m}}+3H\rho_{m}=Q \label{2e1}\end{equation}
\begin{equation}
\dot{\rho}_{\Lambda}+3H(1+\omega_{\Lambda})\rho_{\Lambda}=-Q
\label{2e2}
\end{equation}
where $\rho_{m}$ is energy density of matter and
$\omega_{\Lambda}=\frac{p_{\Lambda}}{\rho_{\Lambda}}$ is equation
of state parameter of dark energy. Following \cite{pp} we take the
interaction term $Q$ as a decay process $Q=\Gamma \rho_{\Lambda}$
with $\Gamma$ being an arbitrary decay rate.  If $\Gamma>0$, dark
energy decays into the dark matter. Here we do not concern with
the details of this decay process and do not answer the question
that where exactly the dark energy is going
to, see for instance \cite{al} and references therein.\\
We specialize to Friedman-Robertson-Walker spacetime which is
given by
\begin{equation}
ds^{2}=-dt^{2}+a^{2}(t)\{\frac{dr^{2}}{1-kr^2}+r^{2}(d\theta^{2}+\sin^{2}d\varphi^{2})\}
\label{2e3}\end{equation} in which $a(t)$ is the scale factor and
the spatial curvature $k=0,-1,1$ corresponds to flat, open and
closed universes.  For this metric and the aforementioned matter
and energy components, the equations (\ref{2b}) can be written as
\begin{equation}
3F(H^2+\frac{k}{a^2})=(\rho_{m}+\rho_{\Lambda})+\frac{1}{2}U\dot{\phi}^2-3H\dot{F}+\frac{1}{2}V
\label{2f}\end{equation}
\begin{equation}
F(3H^2+2\dot{H}+\frac{k}{a^2})=-p_{\Lambda}-\frac{1}{2}U\dot{\phi}^2-\ddot{F}-3H\dot{F}+\frac{1}{2}V
\label{2g}\end{equation} where $H$ is the Hubble parameter defined
by $H\equiv\frac{\dot{a}}{a}$. Note that the field equation of
$\phi$ is not independent of equation (\ref{2b}) and the energy
balances (\ref{2e1}) and (\ref{2e2}).
\subsection{The spatially flat case}
We intend now to obtain the equation of state parameter
$\omega_{\Lambda}$ for a spatially flat spacetime, $k=0$.  We
first choose the Hubble horizon as the IR cutoff, i.e.,
$L=H^{-1}$.  In this case the relation (\ref{03}) takes the form
\begin{equation}
\rho_{\Lambda}=3c^2 M^{2}_{p}H^2 \label{2h}\end{equation} For
mathematical convenience, we shall assume that
$F(\phi)=\phi^{\alpha}$ in which $\alpha$ is a constant parameter.
Following \cite{ban} and \cite{set}, we then restrict our analysis
to the class of solutions for which the scalar field evolves as a
power law of the scale factor $\phi \propto a^n$ with $n$ being a
constant\footnote{In principle, there is no physical reason for
this choice and we find it mathematically convenient.}. Now, by
combining these expressions for $\phi$ and $F(\phi)$ with
equations (\ref{2f}), (\ref{2g}) and (\ref{2h}) and using
(\ref{2e1}), we obtain
\begin{equation}
\omega_{\Lambda}=(1+r)\frac{(\alpha n+2)\frac{\Gamma}{H}-\alpha
n(2\alpha n+3)+VH^{-2}\phi^{-\alpha}-n^2
U\phi^{-\alpha+2}}{3[\alpha n-(\alpha n+2)r]-n^2
U\phi^{-\alpha+2}-VH^{-2}\phi^{-\alpha}} \label{2i}\end{equation}
where $r\equiv\frac{\rho_{m}}{\rho_{\Lambda}}$. This relation
clearly implies that $\omega_{\Lambda}$ is not necessarily a
constant and can take negative values. The latter only sets some
bounds on the numerical values of $\alpha $, $n$ and also possible
dependence of $U$ and $V$ on $\phi$\footnote{It should be pointed
out that every couple ($\alpha$, $n$) does not correspond to a
real solution in this model.  Moreover, one should in general
consider $n({\alpha})$. See, for example, \cite{cappa}.}. Before
making a closer look at (\ref{2i}), let us write the deceleration
parameter $q=-(1+\frac{\dot{H}}{H^2})$. To this end, we combine
(\ref{2g}) and (\ref{2h}) which gives
\begin{equation}
q=\frac{1}{(\alpha n+2)}[1+\alpha n(\alpha n+2)+(3c^2
M^{2}_{p}\omega_{\Lambda}+\frac{1}{2}n^2
U\phi^2-\frac{1}{2}VH^{-2})\phi^{-\alpha}]
\label{2l}\end{equation} Note that $q$ depends on $\Gamma$ and $r$
through the equation of state parameter $\omega_{\Lambda}$.  To
investigate the role of the functions $U(\phi)$ and $V(\phi)$ on
the evolution of the parameters $\omega_{\Lambda}$ and $q$, we
consider two special cases :\\
1) when $n$ is zero the scalar field becomes trivial and takes a
constant configuration.  In this case if we set $V=0$ and $F=(8\pi
G)^{-1}$, the action (\ref{2a}) reduces to the Einstein-Hilbert
action.  Thus, this is the limit of the model to general
relativity. At this limit, the equation (\ref{2i}) reduces to
\begin{equation}
\omega_{\Lambda}=-(1+\frac{1}{r})\frac{\Gamma}{3H}
\label{2i1}\end{equation} which is the result obtained in
\cite{pp}. When there is no interaction, namely $\Gamma=0$, then
$\omega_{\Lambda}=0$ and there is no late time acceleration.  This
dustlike equation of state was the basic problem that leads  Li
\cite{li} to take the future event horizon rather than the Hubble
radius as the IR cutoff. Moreover, the deceleration parameter
(\ref{2l}) also reduces to
\begin{equation}
q=\frac{1}{2}[1-3c^2 M^{2}_{p}(1+\frac{1}{r})\frac{\Gamma}{3H}]
\label{2l1}\end{equation} in which we have used (\ref{2i1}). Thus,
$q=\frac{1}{2}>0$ when $\Gamma=0$ and, as previously stated, there
is no non-interacting limit.\\
2) the parametrization $\alpha=1$ and
$U(\phi)=\frac{\omega}{\phi}$ corresponds to the Brans-Dicke model
with a scalar field potential. In this case, the relation
(\ref{2i}) reduces to
\begin{equation}
\omega_{\Lambda}=(1+r)\frac{(n+2)\frac{\Gamma}{H}-(2n+3)n-n^2
\omega +V(\phi)H^{-2}\phi^{-1}}{3[n-(n+2)r]-n^2
\omega-V(\phi)H^{-2}\phi^{-1}} \label{2j}\end{equation} This
implies that $\omega_{\Lambda}$ can take negative values when
\begin{equation}
(n+2)\frac{\Gamma}{H}+\frac{V(\phi)}{\phi
 H^2}>(2n+3)n+n^{2}\omega
\label{2n}\end{equation}
\begin{equation}
3(n+2)r+n^{2}\omega +\frac{V(\phi)}{\phi H^2}>3n
\label{2o}\end{equation} or the reversed direction for the both
inequalities. These conditions clearly depend on numerical values
of $\Gamma$, $\omega$ and $n$ as well as the potential $V(\phi)$.
When $V(\phi)=0$, the above inequalities set constraints on the
decay rate of the dark energy into the dark matter. For a nonzero
potential, on the other hand, we can consider more solutions of
the field equations in which the set of parameters satisfies the
inequalities (\ref{2n}) and (\ref{2o}). For instance, for a
quartic potential \cite{p1} $V(\phi) \sim \phi^{4}$, the potential
term $\frac{V(\phi)}{\phi H^2}$ increases with time so that at
late times it dominates the conditions (\ref{2n}) and
(\ref{2o})\footnote{We
assume that $n>0$. For $n<0$, one may consider an inverse power law potentials \cite{p2}}.\\
In this parametrization, deceleration parameter (\ref{2l}) reduces
to
\begin{equation}
q=\frac{1}{(n+2)}\{1+n(n+2)+\frac{1}{2}n^2 \omega +3c^2
M^{2}_{p}\omega_{\Lambda}\phi^{-1}-\frac{1}{2}V(\phi)H^{-2}\phi^{-1}\}
\label{2l2}\end{equation}  The signature flip of the deceleration
parameter from positive to negative values depends crucially on
the evolution of the negative term in (\ref{2l2}). At the first
glance, it seems to be important that whether the scalar field
increases or decreases with time, namely that $n>0$ or $n<0$
respectively. However, the both cases may actually happen if the
potential function in the last term on the right hand side of
(\ref{2l2}) has an appropriate functional form. In fact, for a
suitably chosen potential this term may be set to increase with
cosmic time regardless of the sign of $n$.  For instance, for
$n>0$ a power law \cite{p1} and for $n<0$ an inverse power law
\cite{p2} potentials make the negative term of (\ref{2l2}) be an
increasing function of time. In this case, at early times the
potential term is negligible and the deceleration parameter is
effectively positive whereas at late
times it grows and dominates so that $q$ changes its sign.\\
The fact that the both cases $n>0$ and $n<0$ may fit the
deceleration parameter with observations, is an improvement with
respect to the case of zero potential function. In that case,
there is no transition from deceleration to acceleration phase
unless $n<0$ or $\phi$ being a decreasing function of time
\cite{ban}. \\
\subsection{The spatially curved cases}
Although it is a general belief that the universe is spatially
flat, the spatial curvature may still have contribution to the
field equations if the number of e-foldings is not very large. In
fact, recent observations \cite{k0} favor the spatial curvature
and imply that the effect of the latter, though much smaller than
other energy components, can not be completely ruled out.\\
For $k\neq0$, the equations (\ref{2i}) and (\ref{2l}) generalize
to
\begin{equation}
\omega_{\Lambda}=(1+r)\frac{(\alpha n+2)\frac{\Gamma}{H}-\alpha
n(2\alpha n+3)+VH^{-2}\phi^{-\alpha}-n^2
U\phi^{-\alpha+2}-2\Omega_{k}}{3[\alpha n-(\alpha n+2)r]-n^2
U\phi^{-\alpha+2}-VH^{-2}\phi^{-\alpha}+6\Omega_{k}}
\label{2ii}\end{equation}
\begin{equation}
q=\frac{1}{(\alpha n+2)}[1+\alpha n(\alpha n+2)+\Omega_{k}+(3c^2
M^{2}_{p}\omega_{\Lambda}+\frac{1}{2}n^2
U\phi^2-\frac{1}{2}VH^{-2})\phi^{-\alpha}]
\label{2ll}\end{equation} where $\Omega_{k}=\frac{k}{a^{2}H^{2}}$.
To explore the consequences of these equations, let us write them
in the two special cases considered above.  Firstly, in the limit
to general relativity, (\ref{2ii}) and (\ref{2ll}) reduce to
\begin{equation}
\omega_{\Lambda}=-\frac{1}{3}(1+r)\frac{\frac{\Gamma}{H}-\Omega_{k}}{r-\Omega_{k}}
\label{2i2}\end{equation}
\begin{equation}
q=\frac{1}{2}[1+\Omega_{k}+3c^2 M^{2}_{p}\omega_{\Lambda}]
\label{2l22}\end{equation} For $k=0$, these equations reduce to
(\ref{2i1}) and (\ref{2l1}).  Inspection of (\ref{2i2}) reveals
that for a closed universe ($\Omega_{k}>0$) additional conditions
should be satisfied with respect to the flat case.  Specifically,
the conditions $\frac{\Gamma}{H}>\Omega_{k}$ and $r>\Omega_{k}$,
or the reversed direction of the inequality for the both, must
hold in order that $\omega_{\Lambda}$ remains negative.  It is
interesting to note that contrary to the flat case, the relation
(\ref{2i2}) allows a non-interacting limit.  It means that for an
open universe ($\Omega_{k}<0$), $\Gamma=0$ still gives a
consistent result.  For a closed universe, the same is true if
$r<\Omega_{k}$.\\
Secondly, in the Brans-Dicke parametrization, (\ref{2j}) and
(\ref{2l2}) generalize to
\begin{equation}
\omega_{\Lambda}=(1+r)\frac{(n+2)\frac{\Gamma}{H}-(2n+3)n-n^{2}
\omega+V(\phi)H^{-2}\phi^{-1}-2\Omega_{k}}{3[n-(n+2)r]-n^2
\omega-V(\phi)H^{-2}\phi^{-1}-6\Omega_{k}}
\label{2kk}\end{equation}
\begin{equation}
q=\frac{1}{2}\{1+n(n+2)+\frac{1}{2}n^2 \omega +\Omega_{k}+3c^2
M^{2}_{p}\omega_{\Lambda}\phi^{-1}-\frac{1}{2}V(\phi)H^{-2}\phi^{-1}\}
\label{2l222}\end{equation} It seems that introducing the
curvature density in (\ref{2kk}) does not alter seriously our
qualitative picture of the equation of state parameter. However,
we should set further conditions to have $\omega_{\Lambda}<0$. We
should have $(n+2)\frac{\Gamma}{H}+\frac{V(\phi)}{\phi
 H^2}>(2n+3)n+n^{2}\omega+2\Omega_{k}$
together with $3(n+2)r+n^{2}\omega +\frac{V(\phi)}{\phi
H^2}+6\Omega_{k}>3n$ or the reversed direction for the both
inequalities.  In these conditions $\Omega_{k}=+1$ or $-1$ for a
closed and open universe, respectively. Inspection of the
deceleration parameter in both cases reveals that the curvature
density only affect the time of onset of the acceleration.  For a
closed universe the acceleration phase is delayed with respect to
the flat case while for an open universe it starts in an earlier
time.
\section{Conclusions}
We generalized holographic dark energy model to scalar-tensor
theories of gravity.  In our model, matter part consists of dark
matter which is assumed to interact with the holographic dark
energy. To explore the parameters $\omega_{\Lambda}$ and $q$, we
restrict our attention to a specific functional form of $F(\phi)$,
namely $F(\phi)\propto \phi^{\alpha}$.  We have also assumed that
the evolution of the scalar field is related to that of the scale
factor in such a way as $\phi \propto a^{n}$. It is then shown
that the desired behavior for $\omega_{\Lambda}$ and $q$ is
possible for a wide range of the functions $U(\phi)$ and
$V(\phi)$.  Specifically, we have considered two special cases. In
the limit of the model to general relativity the results of
\cite{pp} recovered in the flat case.  In the case of non-flat
spaces, however, we have shown that $\Omega_{k}$ gives possibility
to $\omega_{\Lambda}$ to take negative values even for $\Gamma=0$
for both closed and open universes.\\
We have also considered the limit to the Brans-Dicke model with a
self-interacting scalar field.  Although it is shown that for
$V(\phi)=0$ the equation of state parameter may have negative
values and $q$ can have a signature flip both in flat and curved
spaces, the existence of the potential function improves the
behavior of these parameters. Specifically, it should be
emphasized that the deceleration parameter can change its sign
during cosmic evolution for both $n>0$ and $n<0$, whereas in the
case of zero potential it is only
possible for $n<0$ \cite{ban}.\\
It is shown that holographic energy is not compatible with phantom
energy \cite{bak}. Thus we must impose $\omega \geq -1$. Combining
this requirement with (\ref{2kk}), results in
\begin{equation}
r\geq
\frac{2n^{2}(\omega+1)-(n+2)\frac{\Gamma}{H}+8\Omega_{k}}{(n+2)\frac{\Gamma}{H}+
V(\phi)H^{-2}\phi^{-1}-n^{2}\omega-2(n^{2}+3n+3)-2\Omega_{k}}
\end{equation}
This constraint can be satisfied, for instance, for a quartic
potential at late times.\\\\\\

{\bf Acknowledgment}\\\\
The author would like to thank anonymous referees for their useful
comments.  This work is supported by the Office of Scientific
Research of Shahid Rajaee University under the contract
No.38227-7.

~~~~~~~~~~~~~~~~~~~~~~~~~~~~~~~~~~~~~~~~~~~~~~~~~~~~~~~~~~~~~~~~~~~~~~~~~~~~~~~~~~~~~~~~

\end{document}